\documentclass{icrc2009}

\usepackage{graphicx}   
\usepackage[font=footnotesize]{subfig} 
\usepackage{fixltx2e}
\usepackage{url}

\newcommand{\shorttitle}[1]%
{\markboth{Proceedings of the 31\MakeLowercase{$^{st}$} ICRC, {\L}\'{o}d\'{z} 2009}{#1} }

\begin{document}
\title{Observation of Radio Galaxies and Clusters of Galaxies with VERITAS}

\author{\IEEEauthorblockN{Nicola Galante\IEEEauthorrefmark{1} for the VERITAS Collaboration\IEEEauthorrefmark{2}}
                            \\
\IEEEauthorblockA{\IEEEauthorrefmark{1}Harvard-Smithsonian Center for Astrophysics, USA, ngalante@cfa.harvard.edu\\
\IEEEauthorrefmark{2}see R.A. Ong et al (these proceedings) for a full author list}
}

\shorttitle{N. Galante, Radio Galaxies and Cluster of galaxies with VERITAS}
\maketitle

\begin{abstract}
Radio galaxies are the only non-blazar extragalactic objects detected in the VHE ($E >100$~GeV) band. 
These objects enable the investigation of the main substructures of the AGN, in particular the core, 
the jet and its interaction with the intergalactic environment. Clusters of galaxies, instead, have not been 
detected by $\gamma$-ray observatories. These objects are collections of up to thousands of galaxies and are the 
densest large-scale structures in the universe. Galaxy clusters consist of up to 85\% dark matter, 
that could reveal its presence through self-annihilation and VHE $\gamma$-ray emission. The observation of 
non-thermal diffuse radio emission in galaxy clusters suggests the presence of accelerated particles and high magnetic 
fields that can also produce VHE emission. Results from the VERTIAS observations of radio galaxies and 
galaxy clusters will be presented.
  \end{abstract}

 
\section{Introduction}
Radio galaxies are currently the only non-blazar extragalactic objects detected at very high energies         
(VHE: $E>100$~GeV). In the unified model of Active Galactic Nuclei (AGN), the various classifications of AGN are 
strongly dependent on the orientation of the galaxy toward the observer's line of sight. Blazars are AGN which 
contain relativistic plasma jets aligned within a few degrees to our line of sight, causing the observed radiation to be 
highly Doppler-boosted. For radio galaxies this is generally not the case due to wider jet viewing angles, making 
these objects more challenging to detect at VHE. Nevertheless, the discoveries of VHE emission from the radio 
galaxy M~87 by the HEGRA collaboration~\cite{Goetting2003}, and more recently by VERITAS~\cite{Acciari2008}, 
as well as the discovery of NGC~5128 (Centaurus A) at VHE by the HESS collaboration~\cite{Raue2009} have opened new 
possibilities into investigating non-thermal processes in the large-scale structures of AGN. 

Radio galaxies are preferentially located in clusters when compared to other galaxies of comparable mass~\cite{Prestage1988}. 
Their powerful jets energize the intra-cluster medium through termination shocks 
accompanied by particle acceleration and magnetic field amplification. Large scale AGN jets and clusters of galaxies 
are believed to be potential accelerators of cosmic rays~\cite{Dermer2009}, and are therefore of particular interest 
for the cosmic-ray community.
According to the \mbox{Fanroff \& Riley} (FR) classification, radio galaxies can be classified into two main families~\cite{Fanaroff1974}. 
In FR~I objects, the radio emission peaks in the central core and diminishes along the jets, 
while in FR~II objects the outer part of the jets (lobes) are brighter in the radio than the central core of the galaxy.  
FR~I radio galaxies are in general found at low redshifts and have a typical radio power at 178~MHz less than            
\mbox{1025 W Hz$^{-1}$ sr$^{-1}$}. Above this radio power nearly all radio galaxies are FR~II objects. Further distinction 
between the two classes can be found in their discrete spectral properties, i.e. in their excitation emission lines~\cite{Zirbel1995}. 
An interesting aspect of this classification is the large number of shared features between FR~I 
type radio galaxies and BL-Lac type blazars. A possible unification of the two sub-classes of AGN is 
suggested~\cite{UrryPadovani1995}, in which FR~I radio-galaxies are BL-Lac objects observed at larger jet viewing angles. 
Evidence of synchrotron emission at radio to X-ray energies from the extended structures and the core of 
radio-galaxies is well explained by relativistic electrons moving in a beamed relativistic jet~\cite{Ghisellini1993}.          
A possible mechanism for HE-VHE radiation is the Synchrotron-Self-Compton (SSC) process~\cite{Jones1974}, 
where the optical and UV synchrotron photons are up-scattered by the same relativistic electrons in the jet. 
Predictions have long been established for the $\gamma$-ray emission~\cite{Bloom1996} and frequency dependent 
variability~\cite{Ghisellini1989}. Besides leptonic scenarios, several models also consider a hadronic origin for the 
non-thermal emission in jets. Accelerated protons can initiate electromagnetic cascades or photomeson 
processes~\cite{Mannheim1993}, or directly emit synchrotron radiation with consequent X-ray emission and Inverse Compton (IC) 
up-scattering of photons by secondary electrons~\cite{Aharonian2002}. 

Modelling the jet emission with an homogeneus SSC mechanism may imply typical Lorentz 
factors particularly high $\Gamma > 50$ with consequent high Doppler factors and small beaming angles 
$\theta\simeq 1^\circ$~\cite{Krawczynski2002}. This is in conflict with the unification scheme according to 
which FR I radio galaxies and BL-Lac objects are the same kind of object observed at different 
viewing angle. Moreover, these high values for the Doppler factor are in disagreement with the 
small apparent velocities observed in the sub-parsec region of the TeV BL-Lac objects Mrk~421 
and Mrk~501~\cite{Marscher1999}. These considerations suggest a decelerating flow in the jet with a 
consequent gradient in the Lorentz factor of the accelerated particles and in average smaller values 
for it~\cite{GeorganopoulosKanzanas2003}. The first consequence, in fact, is that the fast upstream 
particles ``see'' the downstream produced seed photons with an amplified energy density because of 
the Doppler boost due to the relative Lorentz factor $\Gamma_\mathrm{rel}$ . The IC process is then favoured requiring 
less extreme values for the Lorentz factor and allowing larger values for the beaming angle. In 
a similar way, a spine-sheath structure in the jet consisting of a faster internal spine surrounded 
by a slower layer has been also suggested for the broadband non-thermal emission of VHE BL-Lac 
objects~\cite{Ghisellini2005}. This model is supported by radio maps which show a limb 
brightened morphology~\cite{Giroletti2004} and can explain the HE-VHE emission observed in 
radio galaxies at larger angles ($\theta_\mathrm{layer} = 1/\Gamma_\mathrm{layer} \simeq 20^\circ$).
The model well describes the emission in 
both radio galaxies and ``classical'' BL-Lac objects. Observation of the VHE component from radio 
galaxies is therefore signiÞcant for the multi-structured jet modeling, as it can be related to the 
external lower structure of the jet itself.

Galaxy clusters are also potentially of strong interest to TeV observations.
Galaxy clusters are collections of hundreds to thousands of galaxies and are the
densest large-scale structure in the universe. They are also the largest ($\sim10^{15}$ solar
mass, $\sim$10~Mpc diameter) gravitationally-bound objects. However, the velocities of
individual galaxies within a cluster often exceed the cluster's escape velocity, implying
the presence of either an an additional attractive force besides gravity or an additional
invisible mass component (i.e. dark matter). Galaxy clusters also contain large amounts
of very hot ($\sim 10^8$~K) X-ray-emitting intergalactic gas. Although the gas is about twice
as massive as the galaxies, it is also not enough to contain the galaxies in the cluster.
Assuming this gas is in hydrostatic equilibrium with the cluster's gravitational field,
the total cluster mass is calculated to be several times larger than the mass of the
galaxies and the hot gas. The mass of a typical cluster consists of $\sim$85\% dark matter, 5\%
galaxies, and $\sim$10\% intergalactic gas. Given the large mass and high-density of dark
matter, galaxy clusters may be detectable in VHE gamma-rays via the self-annihilation
of dark matter particles. Non-thermal diffuse radio emission, e.g. radio halos and relics,
is also observed in galaxy clusters and is direct evidence that galaxy clusters are also
sites of relativistic particle acceleration and high magnetic fields that can produce VHE
$\gamma$-rays \cite{Voelk1996}\cite{Keshet2003}\cite{Gabici2004}\cite{Berrington2004}.
Galaxy clusters are one of the few prominent candidate classes of $\gamma$-ray
sources which have not yet been detected by satellite- (e.g. the Fermi $\gamma$-ray Space Telescope) or
ground-based $\gamma$-ray observatories (e.g. VERITAS). The significance of such
detection for the theoretical understanding of non-thermal phenomena in cosmology
cannot be overstated. For example it could be used to measure the mass of a potential
dark matter particle, as well as give indications of its distribution on cosmological scales.
Such observations could also completely unfold the morphology of non-thermal particles
and fields in a cluster.

\begin{table*}[!tp]
\caption{Observed sources. The five columns represent: the source name; the source type
(RG = radio galaxy; CoG = cluster of galaxies); 
the period of observation; the zenith angle range; the total observation time in hours; the flux upper limit in Crab units
(tbp = to be presented at the conference).}\label{tab}
\centering
\begin{tabular}{cccccc}
\hline
\bf{Source} & \bf{Type} & \bf{Period} & \bf{ZA} & \bf{observation time} & \bf{flux U.L.} \\
\hline
NGC~1275 & RG & 01/09 -- 02/09 & $15^\circ$ -- $30^\circ$ & 8~h & 1\%\\
3C~111 & RG & 10/08 -- 12/08 &$15^\circ$ -- $30^\circ$& 11~h & tbp \\
Coma cluster & CoG & 03/08 -- 05/08 & $9^\circ$ -- $20^\circ$ & 19~h& 3\%\\
\hline
\end{tabular}
\label{default}
\end{table*}%

 \section{VERITAS Observations}
 
 VERITAS (the very energetic radiation imaging telescope array system) is an array of four 12~m aperture imaging 
atmospheric Cherenkov telescopes (IACT)~\cite{weeks2002}~\cite{holder2006}, 
operating at the Whipple Observatory in southern Arizona. By indirect 
detection of showers produced by $\gamma$-rays interacting in the atmosphere, VERITAS can achieve a very large
effective area, e.g. 20000 m$^2$ at 300~GeV. The showers are reconstructed from the images of the Cherenkov light they 
produce, as recorded in the $3.5^\circ$ field of view cameras of at least two of the four telescopes. At
zenith, the energy threshold for 
spectral reconstruction is about 150~GeV, and the energy resolution for individual $\gamma$-rays is 15-20\%. The 68\% 
containment radius for reconstructed $\gamma$-ray directions decreases with increasing energy, ranging from 
$0.1^\circ$ to $0.2^\circ$ . VERITAS can detect a $\gamma$-ray source with 5\% of the Crab Nebula flux in 2.5~hours. 

In this work we present three sources observed by VERITAS, two radio galaxies and a cluster of galaxies.

\subsection{NGC~1275}

NGC~1275 (Perseus A, 3C~84) is a radio galaxy located in the center of the Perseus cluster 
and is one of the most unusual early-type galaxies in the nearby universe (z = 0.018). Its radio 
emission is core dominated, but it presents also strong emission lines. In addition, the emission 
line system shows a double structure, a high velocity and a low velocity system. The puzzling 
nature of NGC 1275 is still under debate and makes it difficult to definitely classify it in a standard 
AGN sub-class. Its spectral energy distribution (SED) extends from radio to $\gamma$-rays, although
at VHE only upper limits have been presented so far~\cite{Perkins2006}. A source 
coincident with NGC~1275 at high confidence has been recently detected at high energy (HE) 
$\gamma$-rays by Fermi~\cite{Abdo2009b} reporting an average flux between August and December 2008 
described by a simple power law in the energy range from 100 MeV to 25 GeV with a photon spectral index $-2.17$.
The source was already present in the Fermi bright sources list~\cite{Abdo2009a} that
reported a preliminary average integral flux and a preliminary average spectral index. 
Fermi result is compatible with a detection by VERITAS by mean of a simple extrapolation.

VERITAS observed the source between January and February 2009 at a zenith angle range
between $15^\circ$ and $30^\circ$. All data taken under bad weather or with technical problems
have been excluded. A total  amount of 8~hours has remained for analysis purposes.

\subsection{3C~111}

The radio galaxy 3C~111 has been suggested as a counterpart for the unidentified EGRET $\gamma$-ray source
3EG~J0416+3650~\cite{Sguera}. Due to the large uncertainty on the EGRET $\gamma$-ray source position,
12 X-ray and radio sources can be found inside the $3\, \sigma$ confinement error box. Nevertheless, the radio
galaxy 3C~111 is among the 12 sources the only object that is known to emit both in radio and X-rays.
The $5.3\,\sigma$ detection reported by EGRET with an average flux above 100~MeV of \mbox{$1.3\times 10^{-7}$ cm$^{-2}$ s$^{-1}$}
with a simple power-law photon index $-2.59$ makes this $\gamma$-ray source interesting for an instrument
like VERITAS. If detected, given its higher angular resolution, VERITAS would be able to definitely identify the $\gamma$-ray emitter
with the underlying object.

Data have been taken during fall 2008 at a zenith angle range between $15^\circ$ and $30^\circ$. All data taken
under bad weather conditions or with technical problems have been discharged. Finally, a total
amount of about 11~hours has remained for analysis purposes.

\subsection{Coma Cluster}

Observation of galaxy clusters is done unavoidly during the observation of many radio galaxies. This is the case for NGC~1275
and M~87 where the Perseus and Virgo clusters respectively are observed during the radio galaxy observation.
However, a dedicated study on the cluster itself has been done only
on the Coma cluster up to now. The Coma cluster is a nearby cluster of galaxies which 
is well studied at all wavelengths \cite{Coma1} \cite{Coma2}. 
It is at a distance of 100~Mpc (z=0.023) and has a mass of $2\times 10^{15}\, M_\odot$. Its X-ray and radio features
suggest the presence of accelerated electrons in the inter-galactic medium emitting non-thermal radiation.
Beside relativistic electrons, there might be a population of highly energetic protons. Both high energy electrons and
protons are known to be able to produce VHE photons. A detection by a sensitive instrument like VERITAS
would considerably improve our understanding of the acceleration processes of the intra-cluster medium.

Data have been recorded between March and May 2008. As for the other sources, data taken with technical problems
or under non-optimal weather conditions have been excluded. A total amount of 19~hours spanning a zenith-angle
range of $9^\circ$ to $20^\circ$ have been used for analysis purposes.

 \section{Data Analysis}
 
The data have been analyzed using the VERITAS standard VEGAS analysis package, with background rejection 
cuts optimized on a Crab-like spectrum. The direction and 
impact point of each shower is reconstructed from the positions and orientations of the shower images in the cameras. 
Candidate $\gamma$-ray showers are required to have images with length and width consistent with expectations from 
simulated showers of the corresponding size and distance from each telescope, allowing a large fraction of the 
background proton showers to be rejected. Proton showers typically have longer and wider images. 

Analysis on the radio galaxy 3C~111 is still in progress and will
be presented at the conference. A preliminary analysis on NGC~1275 reveals no signal and 
the excess events distribution over the sky map of the VERITAS FoV is compatible with a Gaussian
distribution of average 0.
An integral upper limit is calculated at the 99\% 
confidence level using the method described by \cite{Rolke} assuming a Poissonian background and a 20\%
uncertainty on the efficiency. A preliminary result above 250~GeV yields an integral upper limit of 
\mbox{$2.58\times 10^{-12}$ erg cm$^{-2}$ s$^{-1}$}.

The analysis of the Coma cluster showed no evidence for point-source emission within 
the field of view and a preliminary upper limit of $\sim3$\% of the Crab flux is given for a moderately extended 
region centered on the core~\cite{Perkins2008}. 

All results are summarized in Table~\ref{tab}.
 
\section{Conclusions}

During the last observation cycle VERITAS observed two radio galaxies, NGC~1275 and 3C~111.
In addition, VERITAS observed the Coma cluster of galaxies. All observations resulted in a null detection.
A preliminary analysis on the radio galaxy NGC~1275 and on the Coma cluster of galaxies resulted in 
upper limits and a final result will be reported at the conference. 
Analysis on the 3C~111 radio galaxy is still in progress and will be reported at the conference.
 
\section*{aknowledgments}
 
This research is supported by grants from the US Department of  
Energy, the US National Science Foundation, and the Smithsonian  
Institution, by NSERC in Canada, by Science Foundation Ireland, and  
by STFC in the UK. We acknowledge the excellent work of the technical  
support staff at the FLWO and the collaborating institutions in the  
construction and operation of the instrument.

\end{document}